\documentstyle[a4,12pt]{article}
\renewcommand{\baselinestretch}{1.5}

\begin{document}
\parskip=5pt plus 1pt minus 1pt

\begin{flushright}
hep-ph/9804434 \\
{\bf LMU-98-06}
\end{flushright}

\vspace{0.2cm}

\begin{center}
{\Large\bf A Determination of the Weak Phase $\gamma$ \\
from Color-allowed $B^{\pm}_u \rightarrow DK^{\pm}$ Decays}
\end{center}

\vspace{0.3cm}

\begin{center}
{\bf Zhi-zhong Xing} \footnote{Electronic address: 
xing@hep.physik.uni-muenchen.de} \\
{\it Sektion Physik, Universit$\ddot{a}$t M$\ddot{u}$nchen,
Theresienstrasse 37A, 
80333 M$\ddot{u}$nchen, Germany}
\end{center}

\vspace{2.5cm}

\begin{abstract}
We show that it is possible to determine the weak phase 
$\gamma \equiv \arg (- V^*_{ub} V_{ud} / V^*_{cb} V_{cd})$ 
of the Cabibbo-Kobayashi-Maskawa flavor mixing matrix only
from the measurement of the color-allowed $B^{\pm}_u
\rightarrow D K^{\pm}$ decay rates. The uncertainty of
this method, arising mainly from the factorization
approximation for two tree-level spectator quark
transitions, may be well controlled.
\end{abstract}

\vspace{1cm}
\begin{center}
PACS number(s): 12.15.Hh, 12.15.Ji, 13.25.Hw, 14.40.Nd
\end{center}

\newpage

The first observation of the Cabibbo-suppressed but color-allowed
transitions $B^+_u \rightarrow \bar{D}^0 K^+$ and $B^-_u \rightarrow
D^0 K^-$, which involve the quark subprocesses $b\rightarrow c 
\bar{u} s$ and $\bar{b} \rightarrow \bar{c} u \bar{s}$ respectively,
has been reported by the CLEO Collaboration \cite{CLEO}. 
These two decay modes, together with $B^+_u \rightarrow D^0 K^+$,
$B^+_u \rightarrow D_{1,2} K^+$ and their charge-conjugate 
counterparts 
\footnote{where $D_1$ and $D_2$ denote the $CP$-even and $CP$-odd
states of $D^0$ and $\bar{D}^0$ mesons, respectively.},
can in principle be used to determine the weak phase
$\gamma \equiv \arg (- V^*_{ub}V_{ud} / V^*_{cb} V_{cd})$ of the
Cabibbo-Kobayashi-Maskawa (CKM) matrix \cite{GW,GL}
\footnote{Possible new physics in $D^0$-$\bar{D}^0$ mixing could
affect those methods proposed in Refs. \cite{GW,GL}. See
Ref. \cite{Xing96} for some detailed discussions.}.
In practice, however, such a method may have
an essential problem, arising from the difficulty in measuring
the color-suppressed transitions $B^+_u \rightarrow D^0 K^+$
and $B^-_u \rightarrow \bar{D}^0 K^-$. To get around this
problem, Atwood, Dunietz and Soni have proposed the measurement of
$B^+_u \rightarrow D^0K^+ \rightarrow (K^-\pi^+)^{~}_{D^0} K^+$,
$B^+_u\rightarrow \bar{D}^0K^+ \rightarrow (K^-\pi^+)_{\bar{D}^0}^{~}
K^+$ and their charge-conjugate channels for the extraction of
$\gamma$ \cite{Atwood}. Furthermore Soffer has shown how to 
remove some uncertainties from the methods mentioned above and
increase their sensitivity by measuring $CP$-conserving phases
at a $\tau$-charm factory \cite{Soffer}.

The interesting question remains if it is possible to constrain $\gamma$
only from observation of the color-allowed $B^{\pm}_u
\rightarrow D K^{\pm}$ decay rates. Gronau has recently discussed this
possibility by proposing a few promising measurables, such as 
the charge-averaged ratios for $B^{\pm}_u$ decays into 
$D$-meson $CP$ and flavor states \cite{Gronau}:
\begin{equation}
R_i \; \equiv \; 2 ~ \frac{\Gamma (B^+_u \rightarrow D_i K^+) ~ + ~
\Gamma (B^-_u \rightarrow D_i K^-)}{\Gamma (B^+_u \rightarrow 
\bar{D}^0 K^+) ~ + ~ \Gamma (B^-_u \rightarrow D^0 K^-)} \; \; 
\end{equation}
with $i=1$ or 2 (the factor 2 on the right-hand side of Eq. (1)
is just taken to normalize $R_i$ to a value close to one).
It is easy to show that there exist two inequalities,
$\sin^2\gamma \leq R_{1,2}$; and one of the two ratios $R_1$ and $R_2$
must be smaller than one except that the value of $\gamma$ 
happens to lie in a narrow band around $\pi/2$.
Thus the measurement of $R_{1,2}$ may in most cases provide a useful
constraint on $\gamma$.

The purpose of this note is to show that $\gamma$ can indeed be
{\it determined} from $R_1$ and $R_2$ through the formula
\begin{equation}
\cos\gamma \; =\; \frac{\kappa}{2} ~ \frac{R_1 - R_2}{R_1 + R_2 - 2} \; \; ,
\end{equation}
where $\kappa \approx 0.077$ is a coefficient related to the 
ratio of the color-suppressed and color-allowed decay amplitudes:
\begin{eqnarray}
r & \equiv & \left | \frac{A (B^+_u\rightarrow D^0 K^+)}
{A (B^+_u \rightarrow \bar{D}^0 K^+)} \right | \nonumber \\
& = & \left | \frac{A (B^-_u \rightarrow \bar{D}^0 K^-)}
{A (B^-_u \rightarrow D^0 K^-)} \right | \; \; .
\end{eqnarray}
The value of $\kappa$ is obtained by means of the isospin symmetry
and the factorization approximation. Since the factorization
approximation made here involves only tree-level spectator quark 
diagrams, the uncertainty associated with $\kappa$ may
be well controlled. In fact, the reported branching ratio
${\cal B}(B^+_u \rightarrow \bar{D}^0 K^+) = (0.257 \pm 0.065 \pm 0.032)
\times 10^{-3}$ \cite{CLEO}
is in good agreement with the prediction from the
factorization scheme \cite{NS}.
Therefore the measurement of the color-allowed
$B^{\pm}_u \rightarrow D K^{\pm}$ decay rates should allow a 
determination of the weak phase $\gamma$ from Eq. (2)
to an acceptable degree of accuracy in the near future.

To derive the formula in Eq. (2) and calculate the parameter $\kappa$,
we begin with 
the effective weak Hamiltonians responsible for $b\rightarrow c \bar{u}
s$, $b\rightarrow u \bar{c} s$ and their charge-conjugate transitions:
\begin{eqnarray}
{\cal H}^{(1)}_{\rm eff} & = & \frac{G_F}{\sqrt{2}} V_{cb} V^*_{us}
\left [ c_1 (\bar{s} u)^{~}_{V-A} (\bar{c} b)^{~}_{V-A}
+ c_2 (\bar{c} u)^{~}_{V-A} (\bar{s} b)^{~}_{V-A} \right ]
~ + ~ {\rm h.c.} \; , \nonumber \\
{\cal H}^{(2)}_{\rm eff} & = &  
\frac{G_F}{\sqrt{2}} V_{ub} V^*_{cs}
\left [ c_1 (\bar{s} c)^{~}_{V-A} (\bar{u} b)^{~}_{V-A}
+ c_2 (\bar{u} c)^{~}_{V-A} (\bar{s} b)^{~}_{V-A} \right ]
~ + ~ {\rm h.c.} \; ,
\end{eqnarray}
where $c_1$ and $c_2$ are QCD correction coefficients. To calculate
the hadronic matrix elements $\langle DK |{\cal H}_{\rm eff}|B \rangle$,
one has to make some approximations. Here we use the factorization
approximation, which factorizes each four-quark operator matrix 
element into a product of two current matrix elements. Wherever there
is a color mismatch between the quark operator and the external state
in this approach, a phenomenological parameter $\xi$ is introduced
and the factorized matrix element becomes related to one of the
following two coefficients \cite{BSW}:
\begin{equation}
a_1 \; \equiv \; c_1 + \xi c_2 \; , ~~~~~~~~
a_2 \; \equiv \; c_2 + \xi c_1 \; .
\end{equation}
The color-suppression factor $\xi$ is naively expected to be $1/3$,
corresponding to exact vacuum saturation. Both $a_1$ and 
$a_2$ have been determined from experimental data \cite{Browder}.
The explicit
knowledge of $\xi$, $c_1$ and $c_2$ will be irrelevant in our
subsequent analysis.

When applying the effective Hamiltonians and the factorization
approximation to $B^{\pm}_u \rightarrow D K^{\pm}$ decays, 
one has to take possible final-state interactions into account.
Since both $D$ and $K$ are isospin
$1/2$ particles, the state $DK$ can be either $I=0$ or 
$I=1$. An isospin analysis made by Deshpande and Dib \cite{DD} 
gives 
\footnote{The factorized matrix element 
$\langle \bar{D}^0 K^-|(\bar{s} c)^{~}_{V-A}|0\rangle \langle 
0 |(\bar{u} b)^{~}_{V-A}|B^-_u\rangle$, corresponding to an
annihilation process, is expected to be formfactor-suppressed 
\cite{Xing,Rosner} and has been neglected here.}
\begin{eqnarray}
A (B^-_u \rightarrow \bar{D}^0 K^-) & = & \frac{G_F}{\sqrt{2}}
~ (V_{ub} V^*_{cs}) ~ \left (\frac{X}{2} e^{i\phi_0} +
\frac{X}{2} e^{i\phi_1} \right ) \; , \nonumber \\
A (B^-_u \rightarrow D^0 K^-) & = & \frac{G_F}{\sqrt{2}}
~ (V_{cb} V^*_{us}) ~ \left (X + Y \right ) e^{i\phi_1} \; ,
\end{eqnarray}
where $\phi_0$ and $\phi_1$ are the strong phase parameters,
$X$ and $Y$ are the factorized hadronic matrix elements:
\begin{eqnarray}
X & = & a_2 ~ \langle \bar{D}^0 |(\bar{u}c)^{~}_{V-A}|0\rangle
\langle K^-|(\bar{s}b)^{~}_{V-A}|B^-_u\rangle \; \nonumber \\
Y & = & a_1 ~ \langle K^- |(\bar{s}u)^{~}_{V-A}|0\rangle
\langle D^0|(\bar{c}b)^{~}_{V-A}|B^-_u\rangle \; .
\end{eqnarray}
Clearly the weak phase difference between $A(B^-_u\rightarrow
\bar{D}^0 K^-)$ and $A(B^-_u\rightarrow D^0K^-)$ amounts to
$-\gamma$ to an excellent degree of accuracy in the standard 
model \cite{Xing96}.
The strong phase difference between these two amplitudes, denoted
by $\delta$ as the notation
in Ref. \cite{Gronau}, is equal to $(\phi_0 - \phi_1)/2$.
By convention, we take
$(\phi_1 - \phi_0) \in [-\pi, +\pi]$. Then $\cos\delta \geq 0$ holds.

In Ref. \cite{Gronau} the ratio $r$ and the strong phase
difference $\delta$ are formally taken 
as two independent parameters. In our factorization approach, however,
$r$ depends on $\delta$ through the relationship
$r = \kappa \cos\delta$ with 
\begin{equation}
\kappa \; = \; \left | \frac{V_{ub} V^*_{cs}}{V_{cb} V^*_{us}} 
\right | ~ \left | \frac{X}{X + Y} \right | \; \; ,
\end{equation}
derived from Eqs. (3) and (6).
The point is simply  that the amplitude of $B^-_u\rightarrow
\bar{D}^0 K^-$, after its CKM coefficient is factored out, 
contains two isospin components with equal magnitude
but different phases. 
The result of $R_{1,2}$ obtained by Gronau \cite{Gronau} can be 
reproduced as follows:
\begin{equation}
R_{1,2} \; = \; 1 ~ + ~ r^2 ~ \pm ~ 2r ~ \cos\delta ~ \cos\gamma \; .
\end{equation}
By use of $r = \kappa \cos\delta$, we arrive at
\begin{equation}
R_{1,2} \; = \; 1 ~ + ~ \kappa \cos^2\delta 
\left (\kappa \pm 2 \cos\gamma \right ) \; .
\end{equation}
It is easy to obtain two inequalities,
$\sin^2\gamma \leq R_{1,2}$, from either Eq. (9) or Eq. (10). 
Note that there exists a narrow band around 
$\gamma =\pi/2$, which makes both $R_1$ and $R_2$ equal to or larger than
one. With the help of Eq. (10), 
we find that the necessary condition for $R_{1,2} \geq 1$ is
\begin{equation}
- \frac{\kappa}{2} \; \leq \; \cos\gamma \; \leq \;
+ \frac{\kappa}{2} \; .
\end{equation}
Beyond this band of $\gamma$, one of the two ratios $R_1$ and $R_2$
must be smaller than one, thus one may get
the constraint $\sin^2\gamma \leq R_1 <1$ or 
$\sin^2\gamma \leq R_2 <1$.

Note that the weak phase $\gamma$ can indeed be determined from $R_1$
and $R_2$, provided the value of $\kappa$ is known. From
Eq. (10), we straightforwardly obtain 
\begin{equation}
\cos\gamma \; =\; \frac{\kappa}{2} ~ \frac{R_1 - R_2}{R_1 + R_2 - 2} \; \; .
\end{equation}
This instructive result has been listed in Eq. (2). 
To evaluate $\kappa$, we express the hadronic matrix elements $X$ and $Y$
in terms of relevant decay constants and formfactors. The ratio $X/Y$
turns out to be
\begin{equation}
\frac{X}{Y} \; =\; \frac{a_2}{a_1} ~ \frac{m^2_B - m^2_K}{m^2_B 
-m^2_D} ~ \frac{f_D}{f_K} ~ \frac{F^{BK}_0(m^2_D)}{F^{BD}_0(m^2_K)} \;\; .
\end{equation}
The main error bar of $X/Y$ comes from the unknown decay constant $f_D$.
Here we typically take $f_D = 220$ MeV. We also input 
$f_K = 160$ MeV, $a_2/a_1 =0.25$ \cite{Browder}, $F^{BK}_0(0) =0.38$,
and $F^{BD}_0(0) =0.69$ \cite{BSW}. 
The uncertainties of the chosen values for $F^{BK}_0(0)$ and $F^{BD}_0(0)$ 
are expected to be within $15\%$ \cite{BallCheng}.
By use of a simple monopole model
for formfactors \cite{BSW}, we obtain $F^{BK}_0(m^2_D) \approx 0.43$
and $F^{BD}_0(m^2_K) \approx 0.70$. Then we arrive at $X/Y \approx 0.24$.
With the input $|(V_{ub} V^*_{cs})
/ (V_{cb} V^*_{us})| = 0.4$ \cite{Drell},
we finally get $\kappa \approx 0.077$. This implies that the value
of $r = \kappa \cos\delta$ is smaller than the naive expectation
$r \approx 0.1$, obtained in Ref. \cite{Gronau} with the assumtion $\delta =0$
and $X/(X+Y) \approx a_2/a_1$.
Since the error associated with the CKM factor is only 
$25\%$ or so \cite{Drell} and those associated with the formfactors
can be partly cancelled in the ratio $X/Y$, 
it should be a conceivable argument that the realistic value of
$\kappa$ cannot be greater or smaller than our present result by a 
factor of two. That is, 
$0.04 \leq \kappa \leq 0.15$ should be a sufficiently  generous range 
of $\kappa$. Even if the uncertainty
associated with the factorization approximation itself is taken into
acount, the possibility of $\kappa \geq 0.2$ would remain extremely small.

A significant deviation of the strong phase difference 
$\delta$ from zero,
implying significant rescattering effects of $DK$ states, may
reduce the magnitude of $r$ further. Considering $DK$ scattering
via a $t$-channel exchange of Regge trajectories, Deshpande and
Dib have made an estimation of the strong phase difference 
$(\phi_1 -\phi_0)$ \cite{DD}. They obtained
$\tan (\phi_1 -\phi_0) \approx -0.14$, equivalent to
$\delta \approx 4^{\circ}$. This value leads to 
$\cos \delta \approx 1$ as an excellent approximation. 
Nevertheless, one should take such a result more qualitative rather than
quantitative, as the method of Regge scattering itself
involves large uncertainties.  
It has been argued
that rescattering effects in decays of a $B$ meson into
lighter hadrons might not be as small as commonly imagined \cite{FSI}.

Taking $\kappa \approx 0.077$, we obtain the narrow band of
$\gamma$ from Eq. (11): $87.8^{\circ} \leq \gamma \leq 92.2^{\circ}$.
In comparison, analyses of current data show that
$\gamma \sim 65^{\circ}$ with a generous 
range $30^{\circ} \leq \gamma \leq 150^{\circ}$ \cite{CKM}. 
For illustration we plot $R_1$ and $R_2$ as functions of $\delta$
in Fig. 1, where three different values of $\gamma$ have been 
typically taken. One can see that $R_1 >1$ and $R_2 <1$ for
cases (a) and (b); but both $R_1 >1$ and $R_2 > 1$ for case (c),
as $\gamma$ is within the narrow band mentioned above. 
It remains unclear that
to what extent the constraint $\sin^2\gamma \leq R_1 < 1$ or 
$\sin^2\gamma \leq R_2 < 1$ will work.
\begin{figure}
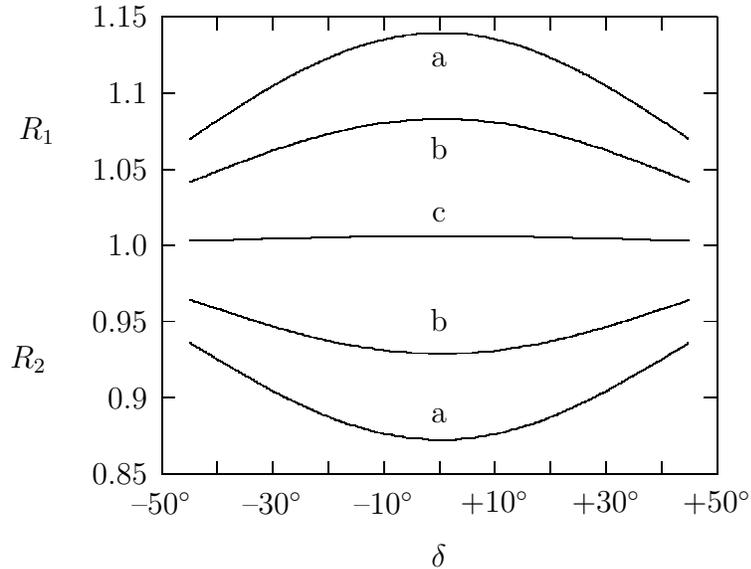

\setlength{\unitlength}{0.240900pt}
\ifx\plotpoint\undefined\newsavebox{\plotpoint}\fi
\sbox{\plotpoint}{\rule[-0.175pt]{0.350pt}{0.350pt}}%

\caption{Illustrative plot for $R_{1,2}$, where
$\kappa =0.077$ and $\gamma=$ (a) $30^{\circ}$;
(b) $60^{\circ}$; and (c) $90^{\circ}$.}
\end{figure}

The idea of extracting the weak phase 
$\gamma$ from $R_{1,2}$ through Eq. (2) or Eq. (12), however,
is valid for all allowed values of $\gamma$. 
The feasibility of this method depends
on a reliable determination of the coefficient $\kappa$.
This should be available in the near future, after $f_D$ is measured from
experiments (or calculated from lattice QCD) and the CKM factor 
$|V_{ub}/V_{cb}|$ is more accurately extracted from charmless $B$ decays.
At least, our present approach can be complementary to that proposed
recently \cite{Gronau} and those suggested
previously \cite{GW,GL,Atwood}. 
It may also confront the nearest data on $B^{\pm}_u\rightarrow
DK^{\pm}$ and give a ballpark number to be expected for $\gamma$,
before the delicate determination of $\gamma$ becomes available
in experriments.

In conclusion, we have shown that it is possible to determine the weak angle
$\gamma$ only from the measurement of the color-allowed $B^{\pm}_u
\rightarrow D K^{\pm}$ decay rates. Such measurements are expected to
carry out
at both $e^+e^-$ $B$-meson factories and high-luminosity hadron machines
in the coming years.

\end{document}